\documentclass[useAMS,usenatbib]{mn2e}
\usepackage{graphicx}
\title[New Brown Dwarfs In  Upper Scorpius]{New brown dwarfs in the south part of the Upper Scorpius Association}
\author[P. Dawson and A. Scholz and T.P.Ray.]{P. Dawson$^{1}$\thanks{E-mail:
dawsonp@tcd.ie (PD); aleks@cp.dias.ie (AS); tr@cp.dias.ie (TR)} and A. Scholz$^{1}$ and T.P. Ray$^{1}$\footnotemark[1]\\
$^{1}$School of Cosmic Physics, Dublin Institute for Advanced Studies, 31 Fitzwilliam Place, 
Dublin 2, Ireland}
\begin{document}

\date{Accepted 2011 xxxxxxx xx. Received 2011 xxxxxxx xx; in original form 2011 April 25}

\pagerange{\pageref{firstpage}--\pageref{lastpage}} \pubyear{2009}

\maketitle

\label{firstpage}

\begin{abstract}
This paper presents the results of a search for brown dwarfs in the Upper Scorpius Association 
using data from the UKIRT Infrared Deep Sky Survey (UKIDSS) Galactic Cluster Survey.   
Candidate young brown dwarfs were first chosen by their position in colour magnitude diagrams 
with further selection based on proper motions to ensure Upper Scorpius membership.   
Proper motions were derived by comparing UKIDSS and 2MASS data. Using that method we identify 19 
new brown dwarfs in the southern part of the association.   In addition there are up to 8 likely 
members with slightly higher dispersion velocity.   The ratio of brown dwarfs to stars was found to be consistent 
with other areas in Upper Scorpius.   It was also found to be similar to other results from 
young clusters with OB associations, and lower than those without, suggesting the 
brown dwarf formation rate may be a function of environment.

\end{abstract}

\begin{keywords}
techniques: photometric -- techniques: brown dwarfs -- open clusters and associations: individual: 
Upper Scorpius -- infrared: stars.
\end{keywords}

\section{Introduction}
The shape of the initial mass function (IMF) is well established and
constrained for stars greater than 0.5$M_{\sun}$. In comparison, 
determining the low mass part of the IMF, particularly in the substellar
regime below $\sim $0.08$M_{\sun}$, has proven more challenging.
In this mass range, the mass function may be affected by turbulent
fragmentation, dynamical interactions, fragmentation of massive disks,
photo-erosion of cores or other processes (see reviews by \citet{whi07,bon07}). 
Hence, in some theoretical scenarios, 
there could be wide variations in the form of the IMF below about 
0.3$M_{\sun}$ depending on the environment. To test these ideas, 
it is essential to carry out surveys for brown dwarfs in diverse 
environments.

Brown dwarfs are difficult to observe because they cool down rapidly with
age. After 1~Myr a very low-mass star near the hydrogen burning boundary 
will have a luminosity only slightly greater than that of the highest 
mass brown dwarf, but after 1~Gyr the brown dwarf will have a luminosity 
an order of magnitude below the star \citep{opp99}. Because of their 
low temperatures and luminosity, searching for them is best done in the 
near infra-red in nearby young open clusters and star forming regions. 
The availability of wide-field near-infrared surveys such as 2MASS and 
UKIDSS thus greatly facilitates searches for substellar objects. 

Most nearby star forming regions have been searched for brown dwarfs
over the past decade (see review by \citet{luh07}). In some clusters, 
the surveys have revealed a population of objects with masses below the 
Deuterium burning limit of 0.015\,$M_{\sun}$ \citep{zap00,lar00}. The ratio
between the number of low-mass stars (0.08-1.0$\,M_{\sun}$) and the number 
of brown dwarfs (0.03-0.08$\,M_{\sun}$) -- an empirical constraint on the IMF --
has been determined to be between 3.3 and 8.5 \citep{and08} and in one region
1.5$\pm$0.3 \citep{sch09}. This might be a first indication for environmental
effects on the IMF.

As of today, the brown dwarf surveys in star forming regions suffer from 
two problems: a) The surveys are incomplete at the low mass end, primarily
due to strong and variable extinction in the molecular clouds. b) Most 
nearby star forming regions are rather similar in their physical
characterics, for example, most of them do not harbour massive stars. 

Here we report on a new brown dwarf survey in a part of the Upper Scorpius
(hereafter UpSco) star forming region. UpSco is a favorable area for
such a project, because it suffers from negligible extinction. At a distance
of $145\pm2$pc \citep{dez99} it is the nearest OB assocation, and it represents 
our best chance of constraining the impact of massive stars on the formation 
of very low mass objects. With an age of about 5~Myr \citep{pre02} UpSco is the 
youngest part of the Scorpius Centaurus Association, i.e. it has one of the better 
combinations of proximity and youth for a successful brown dwarf search. 
UpSco is spread over approximately 250\,deg$^2$ of the sky but wide-field 
surveys now cover a significant portion of this area. 

This paper analyses an infra-red 12\,deg$^2$ survey of part of UpSco, which lies 
roughly within R.A.\ 15h~40m to 16h~20m 
and Dec. -30 to -27 and is generally free of extinction with  $A_{\rm V} < 2.0$ \citep{ard00}.   
Despite its youth (stars later than F type have yet to reach the main sequence) star formation 
appears to have finished in the association within 1~Myr of commencing \citep{pre08,dez99} 
so all members of the association are coeval. In Section 2 details of the survey and the data 
obtained are presented.  Section 3 describes the selection of young brown dwarf candidates 
from photometric and proper motion analysis.   The results of this selection are used to 
analyse the IMF of Upper Scorpius in Section 4. Finally, in Section 5, our conclusions are drawn.

\section[]{Survey And Data Sets}
The United Kingdom Infra-Red Telescope (UKIRT) is currently conducting the 
United Kingdom Infra-red Deep Sky Survey (UKIDSS) - the results of which 
are being made available in a series of releases.   This work used the 8th Data 
Release (DR8Plus).

\begin{figure}
  %\vspace*{174pt}
0  \includegraphics[width=0.45\textwidth]{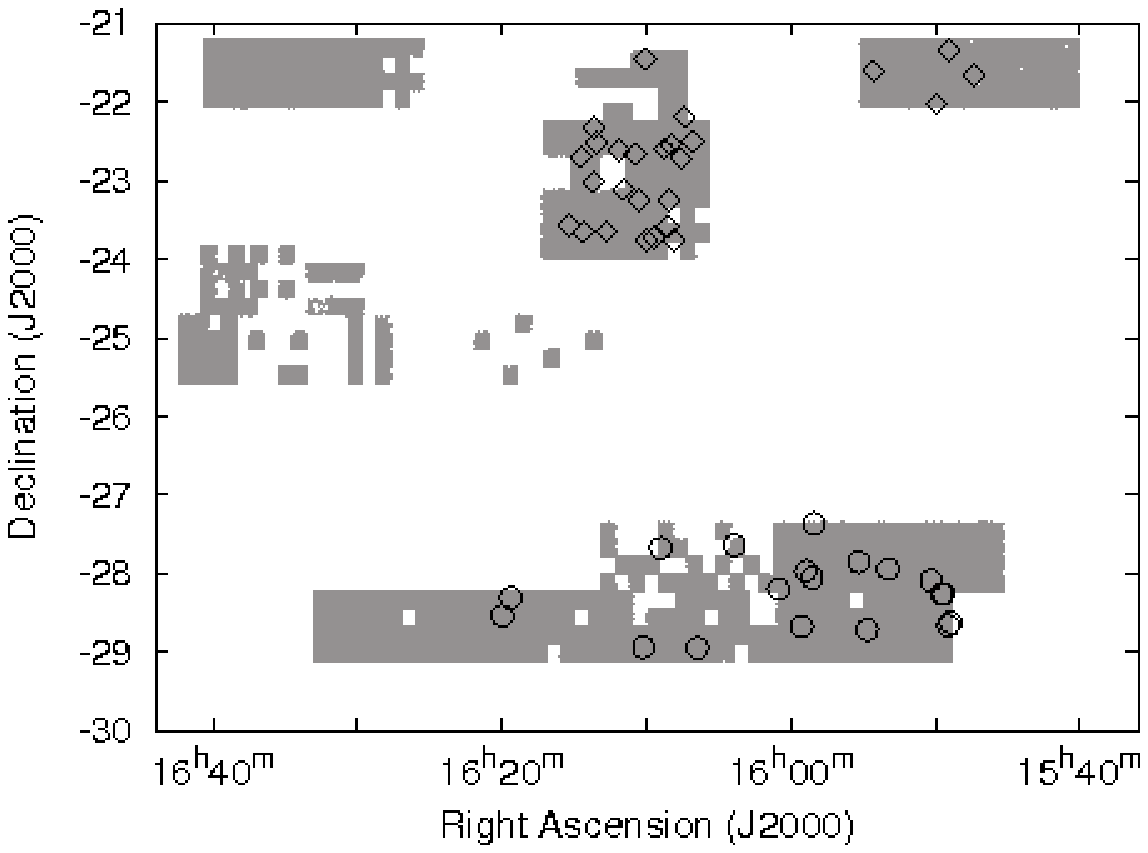}
  %\vspace{1cm}
  \caption{Coverage in Z, Y, J, H and K filters of 28\,deg$^2$ in Upper Scorpius from the UKIDSS GCS 
DR8Plus.   Open circles mark the location of the 19 brown dwarf candidates 
found in this work.    Open diamonds mark the position of spectroscopically 
confirmed brown dwarfs found in other studies \citep{mar04,sle06,lodieu08,lodieu11}.   The search method used 
in this survey of the southernmost area covered by UKIDSS was also applied to 
10.5\,deg$^2$ covered by two areas to the north in order to test its reliability.   
The 26 brown dwarfs shown above were recovered.}
\end{figure}

UKIDSS is made up of several components but the one of interest here is the 
Galactic Cluster Survey (GCS). Described in detail in \citet{lawrence07} the 
GCS is a survey of ten large open star clusters and star forming regions, including 
UpSco. One of the primary goals of the GCS is to conduct a 
census of very low mass brown dwarfs in order to investigate the form 
of the sub-stellar IMF.

The GCS takes infra-red images via five passband filters - Z, Y, J, H and K with 
effective wavelengths of $0.88\mu$m, $1.03\mu$m, $1.25\mu$m, $1.63\mu$m, and 
$2.20\mu$m respectively, and magnitude limits of Z=20.4, Y=20.3, J= 19.5, 
H=18.6 and K=18.6. The instrument used to take the images is the Wide Field 
Camera (WFCAM).  Data collected by the 
WFCAM is subject to an automated process that detects and parameterises objects 
and performs photometric and astrometric calibrations.  The resulting reduced 
image frames and catalogues are then placed in the WFCAM Science Archive (WSA).
The WSA can be interrogated using Structured Query Language (SQL).

A set of five papers provide the reference technical documentation for 
UKIDSS.  \citet{cas07} presents technical details of the WFCAM, 
\citet{hod09} describes the WFCAM photometric system, \citet{ham08} describes 
the WSA and offers instruction on how to extract information from it using 
SQL.  As previously mentioned  \citet{lawrence07} presents the details of the 
different UKIDSS surveys, including the GCS. The fifth paper \citep{irw09} 
% Is this already published??? CHECK
will describe the details of the data reduction pipeline which is run by 
the Cambridge University Astronomical Survey Unit (CASU), but sufficient 
information for an overview of the data reduction pipeline can be gleaned from 
the other four papers and by referring to \citet{dye06} and \citet{war07}\footnotemark[1]\footnotetext[1]{For more details 
of the data reduction process see http://casu.ast.cam.ac.uk/surveys-projects/wfcam/technical.}.

As shown in figure 1, the area in 
UpSco investigated here and surveyed for DR8Plus covers 12\,deg$^2$.   
The data for objects in the target area were obtained via an SQL query 
(see Appendix A for a typical query) to the UKIDSS 
GCS database.   All queries were structured to include only point source 
objects in order to avoid contamination by extended sources (e.g. relatively 
nearby galaxies). Objects in the WSA are given what is known as a discrete 
image classification, with point sources having values between -2 and -1. 
The lines in the query that refer to ``passband''class, e.g. zclass, 
values of between -2 and -1, are designed to 
filter out extended sources.   Note that requiring this value to be between -2 and -1 
in every passband may exclude some sources with very low signal to noise ratios.   As every object with photometric 
characteristics consistent with a brown dwarf had its proper motion assessed, 
in order to check whether it is likely a member of UpSco, each 
query submitted also correlated all objects found in the UKIRT GCS databases 
with those found in 2MASS databases \citep{skrutskie06}.  The 2MASS data is 
used as a first epoch for the purposes of proper motion calculation.

\section{Selection Of Brown Dwarf Candidate Members Of Upper Scorpius Association}

\subsection{Photometry}

\begin{figure*}
%\begin{minipage}{190mm}

\begin{tabular}{|c|c|c|c}

\includegraphics[scale=0.64]{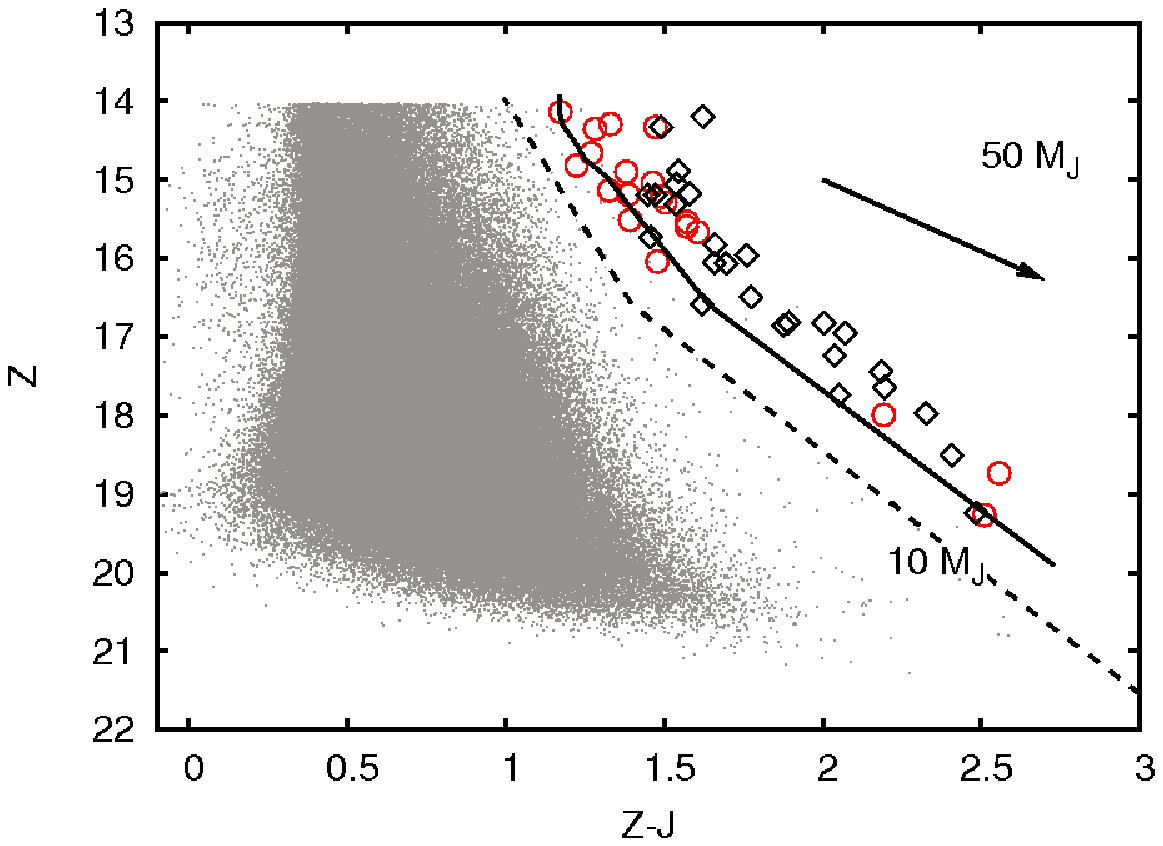}
\includegraphics[scale=0.64]{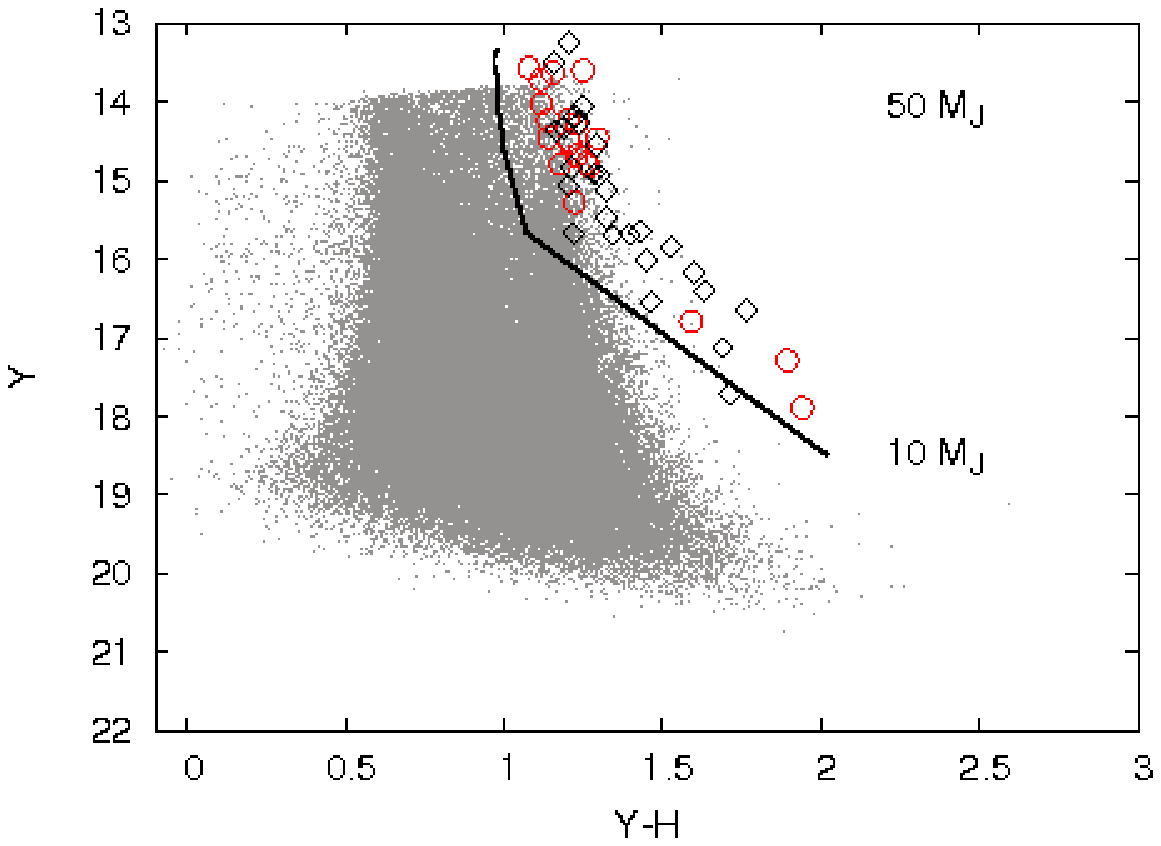}  \\
\includegraphics[scale=0.64]{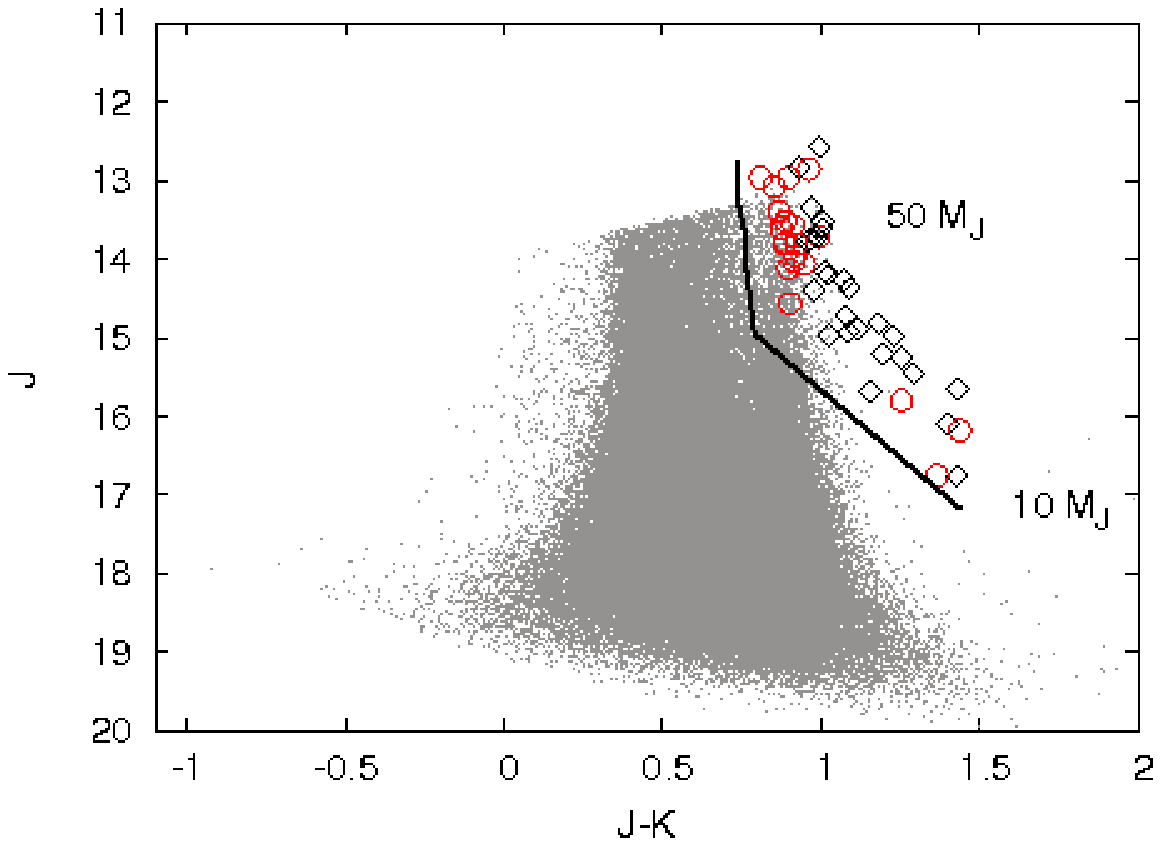} 
\includegraphics[scale=0.64]{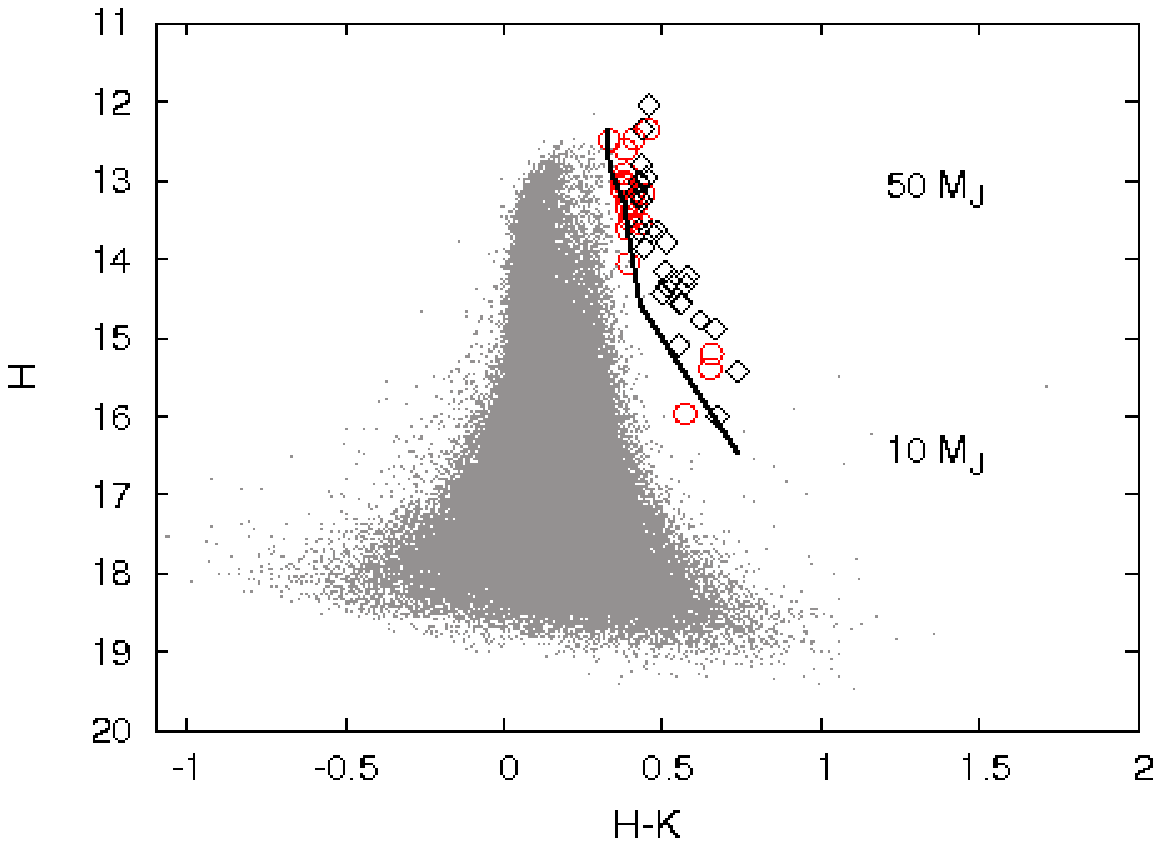}  \\

\end{tabular}
\caption{Four colour magnitude diagrams showing the 19 brown dwarf candidates and the 1 object too faint 
to have its proper motion measured (see text) as red open circles.   
Spectroscopically confirmed brown dwarfs from other studies of UpSco \citep{mar04,sle06,lodieu08,lodieu11} are 
shown as open diamonds and all other objects as small dots.   The 5~Myr DUSTY model \citep{cbah00} isochrone is also 
shown with mass decreasing going down the isochrone as indicated.   As can be seen, some diagrams show a clearer 
demarcation between brown dwarfs and main sequence objects than others.   The (Z-J,Z) diagram was chosen in this work 
for making the photometric selection.   A reddening vector is shown corresponding to the maximum for 
the UpSco region of $A_{\rm V} = 2.0$ noted by \citet{ard00}.   The reddening in the part of UpSco investigated in this 
work is much less than this maximum.   
All objects to the left of the dashed line shown in the (Z-J,Z) diagram were 
rejected because of their colours.}
%\label{fig:CMDs}
%\end{minipage}
\end{figure*}

~
The query shown in Appendix A was submitted to the WSA. The query returned 
282,938 objects and the colour magnitude diagrams shown in 
figure 2 were plotted.  Known brown dwarfs from other studies \citep{mar04,sle06,lodieu08,lodieu11} are shown as open 
diamonds and the 19 brown dwarf candidates found in this study are shown as open circles.   
Theoretical isochrones for 5~Myr old sub-stellar 
objects are also shown over-plotted on the diagrams. These isochrones 
are based on the DUSTY models derived by \citet{cbah00} and obtained from 
both I. Baraffe (private communication) and N. Lodieu (private communication).   The isochrones were 
computed by I. Baraffe using the UKIDSS filter profile.
Reddening caused by extinction shifts the position of objects to the right and down  
on colour magnitude diagrams.   Therefore all brown dwarf candidates should be either on or 
to the right hand side of the isochrones.   The query limited selection to objects 
with magnitudes in Z greater than 14.0.   This choice of a limiting magnitude was motivated 
in part by an examination of colour magnitude diagrams, including those in figure 2, 
which showed that at brighter magnitudes the isochrones for the young brown dwarf/very low 
mass star sequence of objects were no longer sufficiently distinct from other objects on the diagrams.   
Also, the DUSTY models of \citet{cbah00} indicated that this choice would put an upper limit 
on the mass selected of 0.09$M_{\sun}$, massive enough to make sure of including 5~Myr old brown dwarfs 
at 145pc distant.   Note that WFCAM Z is on a Vega system, so it is not directly comparable with the 
SDSS z magnitudes on the AB system.   

%90$M_{\rm J}$

Evident from figure 2 is that some colour magnitude diagrams show 
a much clearer separation between brown dwarfs and main sequence stars than others.   The (Z-J,Z) colour magnitude diagram 
shows the separation clearly and was chosen as the basis for the photometric cut.   Thus to further refine the search, a 
new query was submitted to the WSA eliminating all objects to the left of the 
line from (Z-J,Z) = (1.0, 14.0) through (1.4, 16.6) to (3.0, 21.55) (dashed line in figure 2). 
This query left 51 objects which were examined again in the (Z-J,Z) colour magnitude diagram.   17 of the objects to 
the left of the line (Z-J,Z) = (1.1, 14.0) through (1.1, 14.3), (1.2, 14.9), (1.3, 15.2) to (1.6, 17.0) 
were rejected for being too far from the isochrone on the blue side, leaving 34 photometric candidates.   
Most of the candidates are slightly redder than predicted by the isochrones, which could be due to extinction 
or problems with the isochrones.

\subsection{Proper Motion}

The 34 photometric candidates were then examined to find their 
proper motion.  Proper motions were calculated using the query shown in appendix A.   
The difference in position of the objects in the GCS and 2MASS catalogues is obtained in milliarcseconds.   
This is then divided by the difference in the two epochs, converted from Julian dates to years.   The lines 
in the query that list their results ``as pmRA'' and ``as pmDEC'' perform this task.   
The resulting vector point diagram is shown in figure 3.   The known 
proper motions of UpSco in right ascension and declination are about -11mas/yr and -25mas/yr 
respectively \citep{deb97,pre98}.   Of the 34 candidates, 1 was too faint to be recorded in 2MASS 
leaving 33 candidates with proper motion data calculated.  
The remaining 33 candidates included 6 with proper motions greater than 
the range of figure 3 (Table 2).   These 6 objects might be red or brown dwarfs located much closer to the Sun 
than UpSco.   \citet{deb99} notes that the velocity dispersion in 
UpSco is very small at 1.3km/s, corresponding to about 2mas/yr.   

The greatest contribution to the spread in the proper motions therefore comes from errors in UKIDSS and 2MASS 
measurements.   To assess the errors the original selection of 282,938 objects had their proper motions examined.   
The proper motions were found to have a normal distribution about the origin with a standard deviation of 10.2mas/yr 
in both right ascension and declination.      
Factoring this error back into the \citet{deb99} figure noted above 
showed that a 2$\sigma$ selection circle for UpSco members would have a radius of 20.8mas/yr.   
This error has only a slight dependence on magnitude for objects with 
a magnitude in Z between 14.0 and 17.0.   For the objects fainter than this the standard deviation 
is $\approx$20mas/yr.   There are 3 objects among the final 27 candidates with magnitudes in Z greater than 17.0.

All 27 candidates shown in figure 3 are predominantly centred around the (-11,-25) position.   The 3 candidates 
with magnitudes in Z greater than 17.0 noted above are marked in red.   
There is no clustering of objects around the (0,0) position indicating that the sample is not contaminated by more distant 
objects e.g. AGB stars which have similar surface temperatures and colours to 
brown dwarfs, but much greater intrinsic luminosities.   The 19 candidates within the 2$\sigma$ selection circle 
were then classified as members of UpSco.   These objects so selected 
(Table 1) have the photometric and proper motion characteristics of a 5~Myr old 
brown dwarf member of UpSco.   Given that there are 19 objects within the 2$\sigma$ selection circle statistically 
it is to be expected that possibly 1 of the 8 objects outside is also a brown dwarf member of UpSco.   
However all of the other 8 objects shown in figure 3 are clustered immediately outside the 
2$\sigma$ selection circle and not scattered around the vector point diagram as would be expected for 
random contaminants.   As these 8 objects have the same range of magnitudes as the 19 within the selection 
circle (Table 1) they are not subject to any systemically larger proper motion errors caused by being fainter.   
Thus it is likely that these 8 objects are also members of UpSco with slightly higher dispersion velocity.   
Finally we note that none of the 
brown dwarf candidates listed here have been identified before in previous surveys.  

Given the low contamination in our sample with proper motions, the one object which
is not detected in 2MASS (Table 1) has a high likelihood of being a brown dwarf in UpSco as well
(28/34, i.e. 82\%). 

\subsection{Estimate of Contamination/Completeness}

In order to be certain that the candidates found in this study are in fact brown dwarfs 
spectra should be obtained.   However, figure 3 indicates that there is negligible contamination 
from background stars among the sample.   

As a further test of the method outlined above, it was also used to investigate the 10.5\,deg$^2$ area covered by UKIDSS 
in the two areas shown to the north of figure 1.   The area is also part of UpSco \citep{dez99} 
and therefore any brown dwarf here will share similar photometric and proper motion characteristics 
of those from the area to the south.   After the analysis of the 10.5\,deg$^2$ area was complete, 
49 objects in it were identified as possible brown dwarfs.   All 49 were previously identified 
by \citet{lodieu07,lodieu08} as brown dwarf candidates.   Spectra have been taken of 26 
of the 49 objects \citep{mar04,sle06,lodieu08,lodieu11} and all 26 have been confirmed as 
brown dwarfs.   This result underlines the reliability of the method as a means of discriminating 
brown dwarfs from other objects in UpSco.

In order to estimate completeness levels in all passbands the data from the original 282,938 objects was analysed.   Objects 
were grouped in bins of 0.1 magnitude and examined to see where numbers detected in each bin began to fall.   The 
resulting estimates of 100\% completeness were: Z=18.0, Y=17.4, J=17.2, H=16.2, K=16.1.   In the Z and J passbands these would 
be the expected magnitudes of UpSco member objects in the 0.01 - 0.02$M_{\sun}$ mass range.   The histograms in all five passbands showed that 
completeness fell gradually at these magnitudes and was still at an 80\% level another magnitude deeper.   
Note that the lower mass range achieved in this survey is not limited by the 
completeness of UKIDSS but by that of 2MASS (J$\approx$16) due to the need for 
proper motion measurements.  However, of the 34 photometric candidates, only 
one object was fainter than the sensitivity limit of 2MASS and did not have its proper motion calculated.

\begin{table*}
 %\centering
 \begin{minipage}{170mm}
  \caption{Positions, Z, Y, J, H and K photometry, and proper motion 
of the final 19 objects selected.   Also included are the 8 other 
objects shown in figure 3 and lastly, the object 
too faint to be recorded in 2MASS.   Coordinates are J2000.}
  \begin{tabular}{|c|c|c|c|c|c|c|c|c|c}
  \hline
    Name & R.A. & Dec. & Z Mag. & Y Mag. & J Mag. & H Mag. & K Mag. & $\mu_{\alpha}cos\delta$ & $\mu_{\delta}$ \\
& & & & & & & & mas/yr & mas/yr \\
 \hline
2MASSJ15582376-2721435 & 15:58:23.76 & -27:21:43.7 & 14.35 & 13.72 & 13.07 & 12.60 & 12.22 & -17.20 & -19.05\\
2MASSJ16090168-2740521 & 16:09:01.68 & -27:40:52.3 & 14.33 & 13.60 & 12.86 & 12.35 & 11.89 & -8.75 & -17.46\\
2MASSJ16035573-2738248 & 16:03:55.73 & -27:38:25.1 & 15.19 & 14.48 & 13.80 & 13.28 & 12.88 & -12.64 & -28.09\\
2MASSJ15585793-2758083 & 15:58:57.93 & -27:58:08.5 & 15.13 & 14.45 & 13.81 & 13.31 & 12.93 & -9.46 & -20.79\\
2MASSJ15531698-2756369 & 15:53:16.98 & -27:56:37.2 & 15.53 & 14.68 & 13.96 & 13.45 & 13.04 & -14.61 & -33.48\\
2MASSJ15551960-2751207 & 15:55:19.59 & -27:51:21.0 & 15.59 & 14.77 & 14.02 & 13.51 & 13.11 & -19.02 & -39.21\\
2MASSJ15501958-2805237 & 15:50:19.58 & -28:05:23.9 & 16.04 & 15.27 & 14.56 & 14.05 & 13.66 & -5.02 & -22.51\\
2MASSJ15583403-2803243 & 15:58:34.03 & -28:03:24.5 & 15.21 & 14.46 & 13.72 & 13.17 & 12.73 & -3.61 & -20.20\\
2MASSJ16005265-2812087 & 16:00:52.66 & -28:12:09.0 & 15.03 & 14.27 & 13.57 & 13.04 & 12.66 & -0.38 & -21.28\\
2MASSJ15492909-2815384 & 15:49:29.08 & -28:15:38.6 & 14.29 & 13.62 & 12.96 & 12.47 & 12.06 & -14.25 & -22.20\\
2MASSJ15493660-2815141 & 15:49:36.59 & -28:15:14.3 & 14.66 & 14.02 & 13.39 & 12.91 & 12.52 & -11.87 & -24.36\\
2MASSJ16192399-2818374 & 16:19:23.99 & -28:18:37.5 & 15.29 & 14.52 & 13.78 & 13.31 & 12.90 & -5.33 & -8.51\\
2MASSJ15490803-2839550 & 15:49:08.02 & -28:39:55.2 & 14.82 & 14.23 & 13.60 & 13.09 & 12.72 & -19.86 & -22.31\\
2MASSJ15485777-2837332 & 15:48:57.76 & -28:37:33.4 & 17.99 & 16.79 & 15.80 & 15.20 & 14.55 & -17.24 & -17.39\\
2MASSJ16195827-2832276 & 16:19:58.26 & -28:32:27.8 & 18.74 & 17.29 & 16.18 & 15.39 & 14.74 & -27.34 & -22.75\\
2MASSJ15544486-2843078 & 15:54:44.85 & -28:43:07.9 & 15.51 & 14.79 & 14.12 & 13.61 & 13.22 & -15.80 & -12.59\\
2MASSJ15591513-2840411 & 15:59:15.12 & -28:40:41.3 & 14.13 & 13.56 & 12.96 & 12.49 & 12.15 & -11.40 & -15.19\\
2MASSJ16062870-2856580 & 16:06:28.70 & -28:56:58.2 & 14.90 & 14.21 & 13.52 & 13.01 & 12.63 & -7.91 & -16.46\\
2MASSJ16101316-2856308 & 16:10:13.15 & -28:56:31.0 & 15.67 & 14.81 & 14.06 & 13.54 & 13.11 & -10.36 & -18.99\\
\hline
2MASSJ16051544-2802520 & 16:05:15.44 & -28:02:52.0 & 17.92 & 16.66 & 15.69 & 15.05 & 14.47 & -7.11 & -3.30\\
2MASSJ15552513-2801085 & 15:55:25.11 & -28:01:08.8 & 14.12 & 13.51 & 12.88 & 12.47 & 12.01 & -38.10 & -26.24\\
2MASSJ15502934-2835535 & 15:50:29.32 & -28:35:53.9 & 16.05 & 15.32 & 14.59 & 14.03 & 13.63 & -36.07 & -33.63\\
2MASSJ16190983-2831390 & 16:19:09.82 & -28:31:39.5 & 16.63 & 15.67 & 14.70 & 14.17 & 13.67 & -13.41 & -47.04\\
2MASSJ16035601-2743335 & 16:03:56.00 & -27:43:33.6 & 14.41 & 13.91 & 13.26 & 12.64 & 12.29 & -19.94 & -3.33\\
2MASSJ16145936-2826214 & 16:14:59.37 & -28:26:21.8 & 14.68 & 14.12 & 13.50 & 12.83 & 12.48 & +10.75 & -48.05\\
2MASSJ15551768-2856579 & 15:55:17.70 & -28:56:58.1 & 14.32 & 13.80 & 13.19 & 12.66 & 12.33 & +14.02 & -13.91\\
2MASSJ15504920-2900030 & 15:50:49.19 & -29:00:03.1 & 14.35 & 13.82 & 13.21 & 12.64 & 12.34 & -32.06 & -10.49\\
\hline
UGCSJ154723.32-272907.3 & 15:47:23.33 & -27:29:07.3 & 19.27 & 17.91 & 16.76 & 15.97 & 15.40 & ----- & -----\\
\hline

\end{tabular}
\end{minipage}
\end{table*}

\begin{figure}
  %\vspace*{174pt}
  \includegraphics[width=0.45\textwidth]{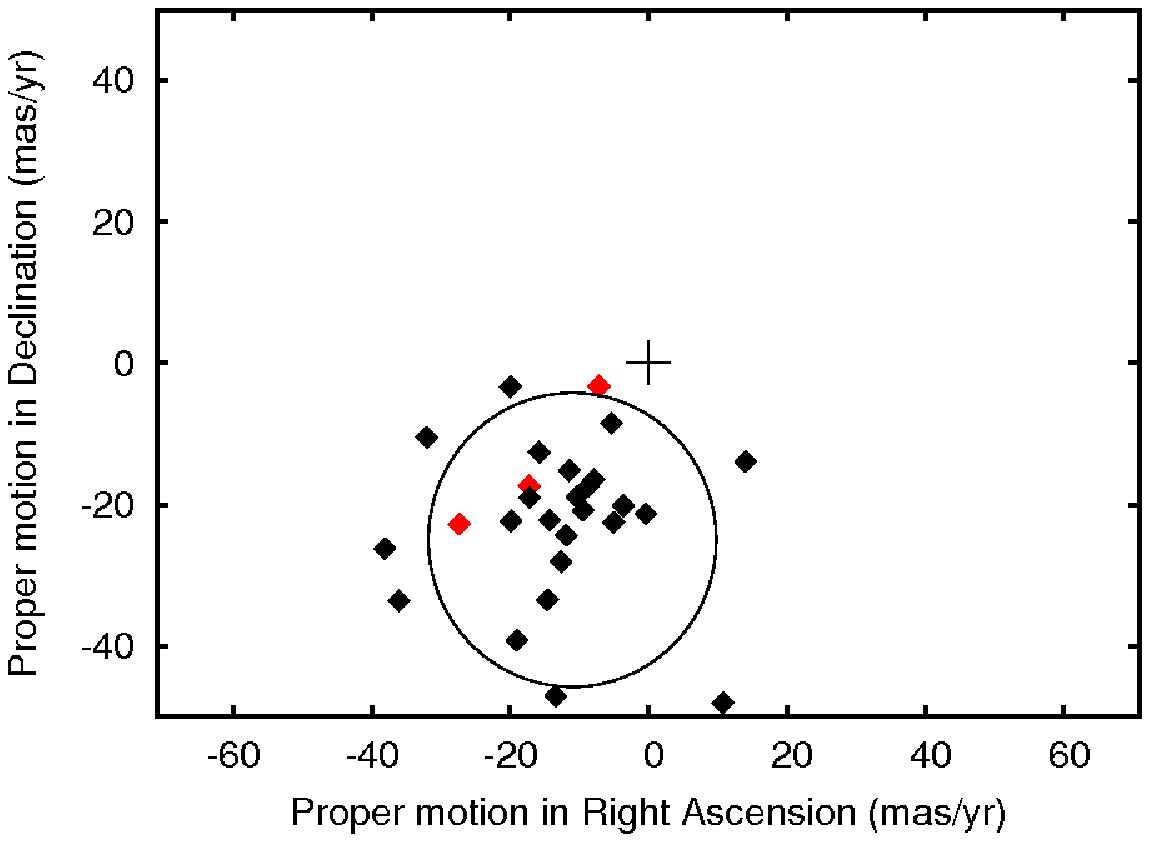}
  %\vspace{1cm}
  \caption{Vector point diagram for 27 candidate brown dwarfs in 
Upper Scorpius.  The 3 objects fainter than Z = 17.0 are marked in red (see text).   
There is an obvious and identifiable cluster 
around (-11,-25), while there is no clustering around the origin, indicating 
that there is very little contamination from background objects in the sample.   
Candidates lying outside the 2$\sigma$ selection circle (see text) centred on (-11,-25) 
were classified as non-members, leaving 19 brown dwarf candidates.   We suspect however that many of the
``outliers'' may well be members.}
\end{figure}

\section{The substellar population in Upper Scorpius}

We re-examined the 19 high probability members listed in Table 1 in the (Z-J,Z) colour magnitude diagram and assigned masses based 
on their Z band magnitude. They cover a mass range from 0.01 to 0.09$M_{\sun}$, 7 are below 
0.03$M_{\sun}$ and 2 of those are below 0.02$M_{\sun}$.   The 100\% completeness limit 
in the Z and J passbands corresponds to a mass of less than 0.02$M_{\sun}$.   At 0.01$M_{\sun}$ the photometric survey 
is still at least 80\% complete.   

\citet{and08} performed a combined analysis of the low-mass IMF in seven star forming regions, not 
including UpSco.   The method used a ratio of stars with masses 0.08 - 1.0$M_{\sun}$ to brown dwarfs 
with masses 0.03 - 0.08$M_{\sun}$ (30 - 80$M_{\rm J}$) from each region to allow for direct comparison.   

In order to follow 
that method a new query was submitted to the WSA to extract UpSco member stars in the mass range 0.09 - 1.0$M_{\sun}$.  
This new query returned 11,041 objects which were then examined in a (Z-J,Z) colour magnitude diagram.   Unlike the brown
dwarfs, the space the UpSco stars occupy in the colour magnitude diagram is not as distinct from the main sequence.   
The isochrones used to guide this selection of UpSco members were the DUSTY isochrone 
used previously and the NexGen isochrone also derived from the models of \citet{cbah00}.   Initially, all objects to the left of a 
line from (0.75,10.32) to (1.17,14.15) were eliminated from consideration.   

The remaining objects then had their proper motion 
examined in the vector point diagram as shown in figure 4.   The UpSco cluster motion is clearly identifiable but is not 
as free from contamination as the brown dwarf selection was.   To estimate contamination, the number of objects contained within 
selection circles of similar size centred on the points (11,25),(25,-11) and (-25,11) were counted (see figure 4).   22 objects were found in all 
three circles so the number of contaminants in the original selection circle was estimated at 7.   This left 
a sample of 37 members of the UpSco within the mass range 0.09 - 1.0$M_{\sun}$ extracted from an 
initial selection of 11,041 objects. Among the brown dwarfs in Table 1, 11 are judged to be in the range 0.03 - 0.08$M_{\sun}$. Note that 
the brown dwarf numbers are extracted from the 19 objects within the 2$\sigma$ selection circle only.   This is to ensure that both star and 
brown dwarf numbers in the ratios are arrived at using the same criteria.   
So the ratio of low-mass stars to brown dwarfs in the selected mass range was found to be 38/11 = 3.5$_{-1.3}^{+2.0}$ (errors are Poissonian).   

The same analysis was then done for 10.5deg$^2$ to the 
north where 49 brown dwarf candidates had been identified.   26 objects were assigned masses of between 0.03 - 0.08$M_{\sun}$.   
Stars in the mass range 0.08 - 1.0$M_{\sun}$ and deemed UpSco members 
because of their photometry and their proper motion, as shown in figure 4, numbered 102.   So the ratio of stars to brown dwarfs 
in those areas was found to be 102/26 = 3.9$_{-0.9}^{+1.4}$, consistent with the south. Overall, 
the combined figures give a ratio of 140/37 = 3.8$_{-0.8}^{+1.1}$ (again, errors are Poissonian).   

These numbers can be compared with the ratios published by \citet[][their Table 1]{and08}. From their sample of 7 clusters, 
we exclude the Pleiades because it is significantly older than all other regions and Mon R2 which has a small population of 
members causing large statistical uncertainties. Only two of the remaining clusters belong to OB associations (ONC, NGC2024), 
and they feature the lowest star/brown dwarf ratios in the sample (3.3 and 3.8). These values are similar to the ones 
derived for the OB association UpSco (3.5 and 3.9, see above). In contrast the clusters without OB associations (Chamaeleon,
Taurus, IC348) in the sample of \citet{and08} have higher star to brown dwarf ratios of 4.0, 6.0 and 8.3, respectively.

Thus, based on the current data it seems possible that the presence of OB stars is related to a low star to brown dwarf
ratio, i.e. a higher abundancy of brown dwarfs. This could be a sign that the radiation field of OB stars favours the
formation of brown dwarfs.   It has been suggested that substellar and planetary-mass
objects can be formed via photoerosion of cores by the ionizing radiation from
an OB star \citep{waz04}.   At face value this would provide an
additional formation channel for brown dwarfs in OB associations, lowering the
star to brown dwarf ratio.   Given the substantial uncertainties in these ratios, this conclusion is certainly preliminary 
and needs to be substantiated by future surveys.

\begin{figure*}
%\begin{minipage}{190mm}

\begin{tabular}{|c|c}

\includegraphics[scale=0.64]{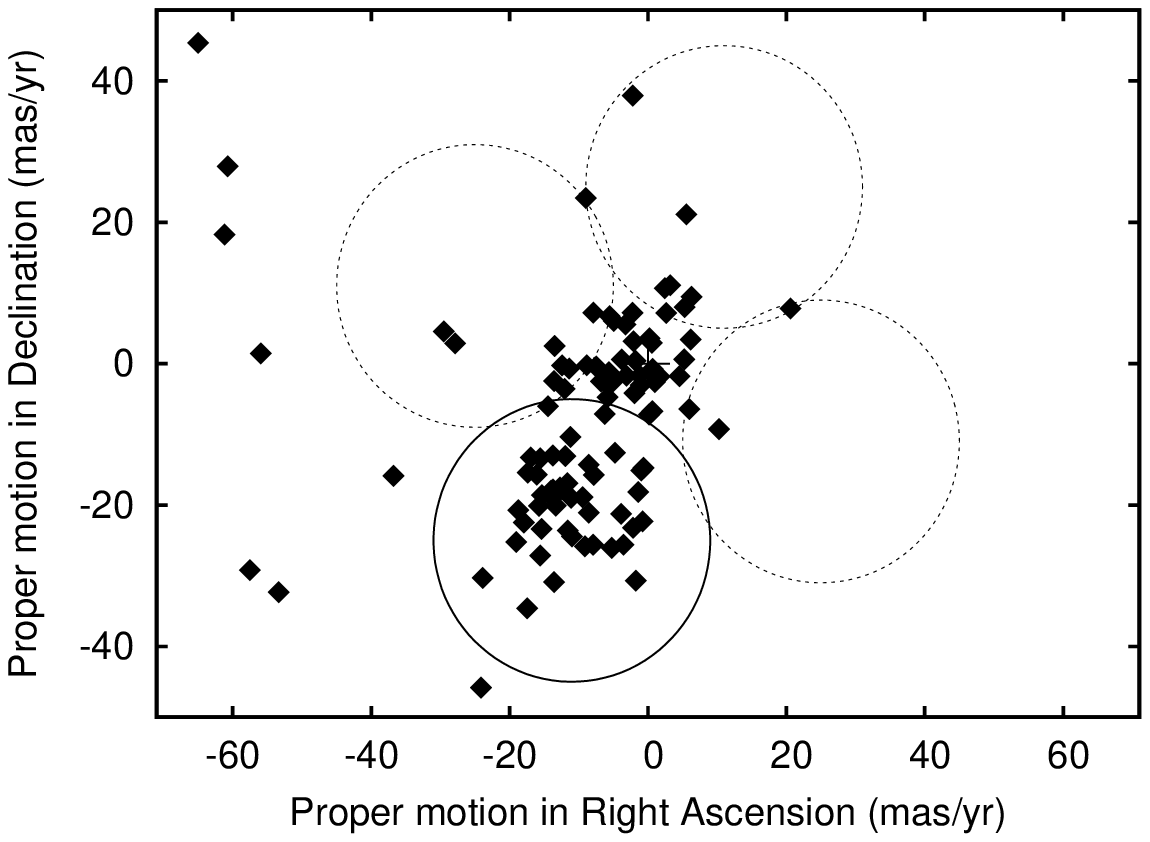} 
\includegraphics[scale=0.64]{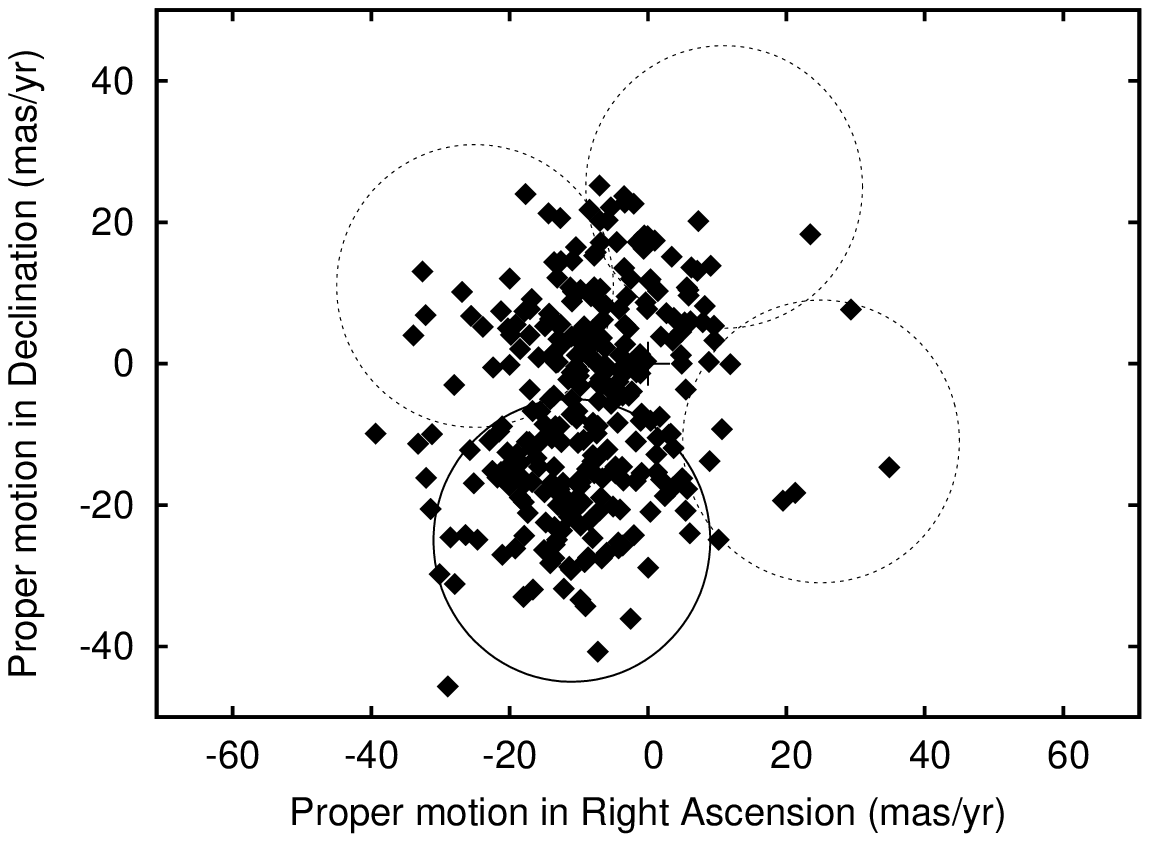}  \\

\end{tabular}
\caption{Vector point diagrams for objects with photometry of stars within mass range 0.08 - 1.0$M_{\sun}$.   As in 
figure 4 a 2$\sigma$ selection circle centred on (-11,-25) is shown.   The diagram on the left shows objects from 
the search area in South UpSco while that on the right is taken from the area to the North.   In both cases the 
Upper Scorpius cluster motion is clearly identifiable.   The dashed circles (see text) were used for estimating contamination.}
%\label{fig:CMDs}
%\end{minipage}
\end{figure*}

\begin{table*}
 %\centering
 \begin{minipage}{170mm}
  \caption{Positions, Z, Y, J, H and K photometry, and proper motion of the 
6 objects outside the range of figure 4.   Coordinates are J2000.}
  \begin{tabular}{|c|c|c|c|c|c|c|c|c|c}
  \hline
    Name & R.A. & Dec. & Z Mag. & Y Mag. & J Mag. & H Mag. & K Mag. & $\mu_{\alpha}cos\delta$ & $\mu_{\delta}$ \\
& & & & & & & & mas/yr & mas/yr \\
\hline
2MASSJ15583064-2802357 & 15:58:30.63 & -28:02:36.3 & 16.53 & 15.75 & 14.99 & 14.47 & 14.03 & -28.27 & -66.30\\
2MASSJ16035915-2806086 & 16:03:58.76 & -28:06:09.6 & 16.78 & 15.82 & 15.02 & 14.46 & 14.00 & -2.19 & -115.84\\
2MASSJ15530915-2828366 & 15:53:09.13 & -28:28:37.3 & 14.81 & 14.18 & 13.51 & 13.00 & 12.63 & -46.20 & -75.62\\
2MASSJ15492508-2843527 & 15:49:24.85 & -28:43:51.6 & 14.15 & 13.49 & 12.84 & 12.39 & 12.02 & -384.23 & +138.80\\
2MASSJ16151681-2907007 & 16:15:16.28 & -29:07:01.3 & 14.10 & 13.40 & 12.86 & 12.23 & 11.96 & +15.06 & -60.37\\
2MASSJ15481934-2748512 & 15:48:19.30 & -27:48:51.3 & 19.77 & 17.56 & 16.73 & 16.15 & 15.83 & -74.00 & -8.34\\
\hline

\end{tabular}
\end{minipage}
\end{table*}

\section{Conclusions}

We have carried out a survey for brown dwarfs in the 5\,Myr old UpSco star forming region based on
photometry and proper motions from a combination of the UKIDSS Galactic Cluster Survey and 2MASS. 19 new
substellar objects with estimated masses between 0.01 and 0.09$M_{\sun}$ are identified. These
objects are located in the southern part of the association which has not been covered by
previous brown dwarf surveys. 8 other objects with slightly higher proper motion have 
also been identified.   These may be UpSco members with slightly higher dispersion velocity 
than the stellar members.   Although spectroscopic confirmation has not been obtained yet, 
the level of contamination appears negligible. 

The ratio of stars to brown dwarfs in the South of UpSco was found to be 
3.5$_{-1.3}^{+2.0}$, in the same range as elsewhere in UpSco.  Comparing with
literature findings, young clusters with OB associations tend to have lower ratios
than clusters without OB stars, which might indicate that brown dwarf formation is 
a function of environment.

\section*{Acknowledgements}

The authors would like to thank Nicolas Lodieu of the Instituto de Astrofisica de 
Canarias and Isabelle Baraffe of the Centre de Recherche Astrophysique de Lyon for 
supplying model data. This work was supported by the Science Foundation Ireland
within the Research Frontiers Programme under grant no. 10/RFP/AST2780.
This publication makes use of data products from the Two Micron 
All Sky Survey, which is a joint project of the University of Massachusetts and the 
Infrared Processing and Analysis Center/California Institute of Technology, funded by 
the National Aeronautics and Space Administration and the National Science Foundation.
We would also like to thank the UKIDSS Team for the excellent database they have made 
available to the community.

\appendix

\section[]{Sample SQL Query}

Shown below is the SQL query submitted to the WSA to find the first set of sources in the Upper Scorpius association.   
The query returned 282,938 rows of data.

\ttfamily
Select\\
g.ra, g.dec, zmypnt, ymjpnt, jmhpnt, hmk\_1pnt, zapermag3, yapermag3, japermag3, hapermag3, k\_1apermag3, 3.6e6*cos(radians(g.dec))*\\(g.ra-T2.ra)/((mj.mjdobs - T2.jdate+2400000.5)/365.25) as\\ pmRA, 3.6e6*(g.dec-T2.dec)/\\((mj.mjdobs - T2.jdate+2400000.5)/365.25) as pmDEC
From\\
gcsmergelog as I, multiframe as mj, (Select t.ra as ra, t.dec as dec, x.slaveobjid as slaveobjid, x.masterobjid as masterobjid, t.j\_m, t.h\_m, t.k\_m, t.jdate From gcssourcextwomass\_psc as x, twomass..twomass\_psc as t Where x.slaveobjid=t.pts\_key And distancemins\\ In (Select Min(distancemins) From gcssourcextwomass\_psc Where \\masterobjid=x.masterobjid)) As T2 Right Outer Join gcssource\\ As g On (g.sourceid=T2.masterobjid)
Where
(g.ra Between 235.0 And 245.0)
And (g.dec Between -30.0 And -27.0)
And zapermag3 > 14.0 And yapermag3 > 11.5 And japermag3 > 12.0 And hapermag3 > 10.0 And k\_1apermag3 > 9.5
And zxi Between -1.0 And +1.0 And yxi Between -1.0 And +1.0 And jxi Between -1.0 And +1.0 And hxi Between -1.0 And +1.0 And k\_1xi Between -1.0 And +1.0
And zeta Between -1.0 And +1.0 And yeta Between -1.0 And +1.0 And jeta Between -1.0 And +1.0 And heta Between -1.0 And +1.0 And k\_1eta Between -1.0 And +1.0
And zclass Between -2 And -1 And yclass Between -2 And -1 And jclass Between -2 And -1 And hclass Between -2 And -1 And k\_1class Between -2 And -1
And (priorsec = 0 Or priorsec = g.framesetid)
And g.framesetid=I.framesetid
And I.jmfid=mj.multiframeid

\bsp

\label{lastpage}

\end{document}